\documentclass[]{phostproc}

\usepackage{lipsum} 

\title{What have we learnt from pulsations of B-type stars?}
\author{Jadwiga Daszy\'nska-Daszkiewicz,$^1$ Przemys{\l}aw Walczak,$^1$ Wojciech Szewczuk,$^1$ Alosha Pamyatnykh$^2$}

\affiliation{$^{1}$Astronomical Institute, University of Wroc{\l}aw, ul. Kopernika 11, 51-622 Wroc{\l}aw, Poland\\
$^{2}$Nicolaus Copernicus Astronomical Center, Polish Academy of Sciences, ul. Bartycka 18, 00-716 Warsaw, Poland}

\shorttitle{What have we learnt about B-type stars from their pulsations?}
\shortauthors{Daszy\'nska-Daszkiewicz \textit{et al.}}

\abs{We review the main results obtained from our seismic studies of B-type main sequence pulsators,
based on the ground-based, MOST, Kepler and  BRITE observations.
Important constraints on stellar opacities, convective overshooting and rotation are derived.
In each studied case, a significant modification of the opacity profile at the depths corresponding to the temperature 
range $\log{T}\in (5.0-5.5)$ is indispensable to explain all pulsational properties.
In particular, a huge amount of opacity (at least 200\%) at the depth of the temperature $\log T = 5.46$ (the "nickel" opacity)
has to be added in early B-type stellar models to account for low frequencies which correspond to high-order g modes.
The values of the overshooting parameter, $\alpha_{\rm ov}$, from our seismic studies is below 0.3.  
In the case of a few stars, the deeper interiors have to rotate
faster to get the g-mode instability in the whole observed range.}

\begin{document}

\maketitle

\section{Introduction}

The main goal of seismic studies is to obtain information on physical processes
and conditions in stellar interiors that determine evolution and the final fate of stars.
The most important ones are differential rotation, mixing processes, convection and opacities.
The power of asteroseismic constraints depends on the richness of the oscillation spectrum
and on the uniqueness of mode identification for individual frequency peaks.

High-precision space observations from MOST, CoRoT, Kepler and BRITE have revolutionized our knowledge about many pulsators.
For example, it appeared that in most (if not all) B-type pulsating stars both pressure (p) and gravity (g) modes are observed, i.e.,
there are hybrid pulsators of  $\beta$ Cep/SPB or SPB/$\beta$ Cep type \citep[e.g.,][]{Degroote2009, Balona2011, Balona2015}.
The $\beta$ Cephei type refers to pulsations in low-order p/g modes which are typical for early B-type stars.
The SPB type (Slowly Pulsating B-type stars) corresponds to g modes of high radial orders which dominate in mid- to late-B type stars.

The above discovery has opened up a possibility of getting more stringent constraints on parameters of the model and theory
because a simultaneous excitation of p and g modes offers probing stellar regions sensitive to various physical processes.
However, the presence of high-order g modes in early B-type stars ($M\gtrsim 8M_\odot$)
and p modes in mid- to late-B type stars ($M\lesssim 8M_\odot$) is a challenging fact still waiting for explanation
because these modes are stable in all standard opacity models \citep[e.g.,][]{Pamyatnykh2004, JDD2017}.

B-type stars play an active role in evolution of the chemical composition and the structure of galaxies. Moreover, those
with masses greater than about 8$M_\odot$ end up as core-collapse supernovae and are primary objects responsible
for the Universe enrichment in heavy elements. 
They are also ones of the main source of ultraviolet radiation and owing to their high luminosities they are visible out 
to large distances.

On the other hand, B-type main sequence stars are relatively simple objects for evolutionary modelling.
Firstly, the energy transport by convection in their envelopes can be neglected with a high accuracy. Secondly, the effect of mass loss 
is very week because their masses are below 16$M_\odot$. Thus, we get rid of two additional free parameters 
describing these phenomena.

Here, we present our results of seismic analysis of B-type main sequence pulsators.
In particular, we summarize in the subsequent sections what we have learnt
about stellar opacities, overshooting from the convective core and on internal rotation from these studies.

\section{Constraints on opacity data}

Opacity data are ones of main ingredients in stellar modelling. In the case of B-type main sequence stars, the shape 
of the opacity profile determines also conditions for heat-driven pulsations. It is now well recognized that, the local maximum 
around the temperature $T=200~000$\,K \citep[e.g.,][]{Moskalik1992, Dziembowski1993, Gautschy1993} is responsible 
for exciting pulsations in these stars. This local maximum
is called the $Z-$bump and it is caused by the large number of absorption lines of iron-group ions 
\citep{Iglesias1991, Iglesias1992, Seaton1993}. There are currently three most used databases of stellar opacities: 
OPAL \citep{Iglesias1996}, OP \citep{Seaton2005} and, the most recent, OPLIB \citep{Colgan2015, Colgan2016}.
Despite the fact, that the discovery of the $Z-$bump explained many puzzles and was a big step forward in astrophysics, 
there are still uncertainties and suggestions that stellar opacities are underestimated at some temperatures 
\citep[e.g.,][]{Bailey2015, Blancard2016}.

Here, we present constraints on opacities derived from seismic studies of B-type main sequence stars.
To this end, we constructed seismic models that fit the well identified frequencies (p-mode dominated B-type stars)
or that reproduce the sequence of period spacing (g-mode dominated B-type stars). 
Besides, we always try to account for mode instability.
In some cases we make use of another seismic probe, the parameter $f$, that is the amplitude of radiative flux variations at the level of the photosphere.
In models of B-type pulsators, the theoretical values of $f$ are very sensitive to opacities.
If the empirical counterparts for a given mode can be obtained, then very strong constraints
on the mean opacity profile may be gained. The empirical values of $f$ are derived from photometric and spectroscopic pulsational amplitudes
with the input from model atmospheres \citep{JDD2003, JDD2005}. In our analysis, we relied on Kurucz LTE models \citep{Kurucz2004}
and on TLUSTY NLTE models \citep{Lanz2007}.

\subsection{P-mode dominated pulsators}

\begin{figure*}[ht]
	\centering
	\includegraphics[width=\linewidth]{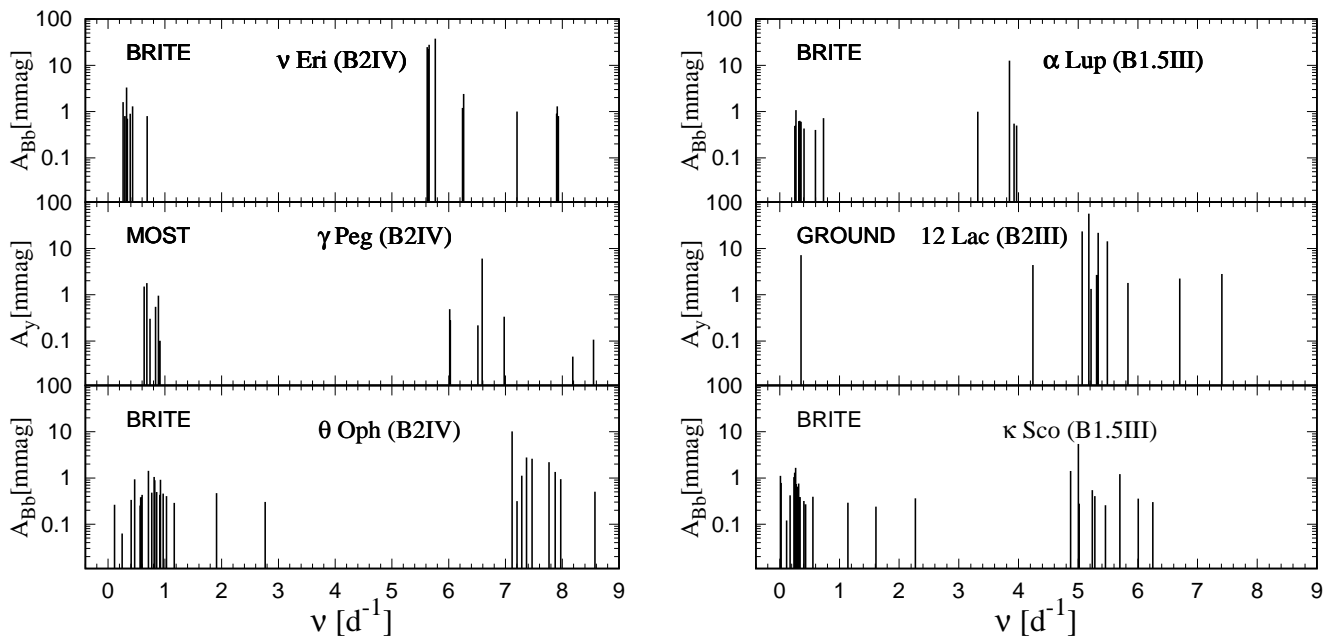}
	\caption{Oscillation spectra of six early B-type stars obtained mostly from space data. All of them have a firm mode identification for several frequency peaks.}
	\label{Fig1}
\end{figure*}

In Fig.\,1, we show the oscillation spectra of studied early B-type stars which are dominated by pulsations in p modes,
i.e., the amplitudes of the p modes are much larger than those of g modes.
As one can see, each star has also some amount of peaks in the low-frequency range, which can correspond only to high-order g modes. 
These low-frequency modes are quite troublesome as their discovery proves some missing components or erroneous assumptions 
in our evolutionary calculations. If we adopt the standard opacity data then no model gives instability in the observed range
of high-order g modes. 

An example is given in Fig.\,2 where  we plot the normalized instability parameter, $\eta$, as a function 
of the mode frequency for the three models suitable for $\nu$ Eridani. The observed frequencies are marked as vertical lines.
These models were computed with the Warsaw-New Jersey evolutionary code \citep{Pamyatnykh1999} and the nonadiabatic code
for linear pulsations of \cite{Dziembowski1977}. Here, the effects of rotation on pulsation were ignored. 
Three opacity data sets were adopted: OPAL, OP and OPLIB. The unstable (excited) modes have positive values of $\eta$.
There are shown modes with the spherical harmonic degree up to $\ell=2$. The effective temperature, $\log T_{\rm eff}$, 
and luminosity, $\log L/L_\odot$, of these models are consistent with the observed error box determined with
the parallax $\pi=4.83(19)$ mas \citep{vanLeeuwen2007}: $\log T_{\rm eff}=4.346(14)$, $\log L/L_\odot=3.886(44)\pm0.044$.
All models have a mass $M=9.5M_{\odot}$ and metallicity $Z=0.015$. The values of ($\log T_{\rm eff},~\log L/L_\odot$) differ
only slightly between the models and their approximate values are: $\log T_{\rm eff}\approx 4.343$, $\log L/L_\odot\approx 3.92$.
All models reproduce the radial fundamental mode $\ell=0,~p_1$ and the centroid of the dipole mode $\ell=1,~g_1$ corresponding 
to the frequencies $\nu=5.76326$ d$^{-1}$ and $\nu=5.63725$ d$^{-1}$, respectively. To adjust the dipole mode frequency 
a small amount of core overshooting was needed, $\alpha_{\rm ov}\in (0.07-0.09)$.
The value of the overshooting parameter $\alpha_{\rm ov}$, in the units of the local pressure scale height, gives a size 
of the layer affected by overshooting from the convective core. 

The widest instability in the p mode range is in the OPLIB model, reaching $\nu\approx 8$ d$^{-1}$,
whereas for the OP model we get the highest values of $\eta$ in the range of high--order g modes.
However, in the allowed range of parameters, there is no model that can account for the whole observed range of frequencies
detected in the $\nu$ Eri light variations. In order to change excitation conditions, we modified the opacity profile, $\kappa(T)$,
according to the formula
$$\kappa (T)=\kappa_0(T) \left[1+\sum_{i=1}^N b_i \cdot\exp\left( -\frac{(\log T-\log T_{0,i})^2}{a_i^2}\right) \right],$$
where $\kappa_0(T)$ is the unchanged opacity profile and $(a,~b,~T_0)$ are parameters of a Gaussian describing
the width, height and position of the maximum, respectively. Thus, for a fixed depth, $T_0$, we enhance or reduce
the opacity by changing $(a,~b)$. The value of $b$ gives approximately the percentage by which the opacity must be changed.
Then, to reproduce the observed frequency ranges as well as the values of some frequency peaks, we introduced small corrections
to $\kappa(T)$ with the following steps $\Delta a=0.001$, $\Delta b=0.05$ and $\Delta T_0=0.005$ in the range $\log T\in (5.0 - 5.5)$.

For each opacity database, we found a plenty of models that reproduced these observed pulsational  features of $\nu$ Eri.
To reduce the number of solutions we employed the nonadiabatic  parameter $f$ whose empirical value was determinable 
for the radial fundamental mode.  
\begin{figure}
	\centering
\includegraphics[width=0.95\linewidth]{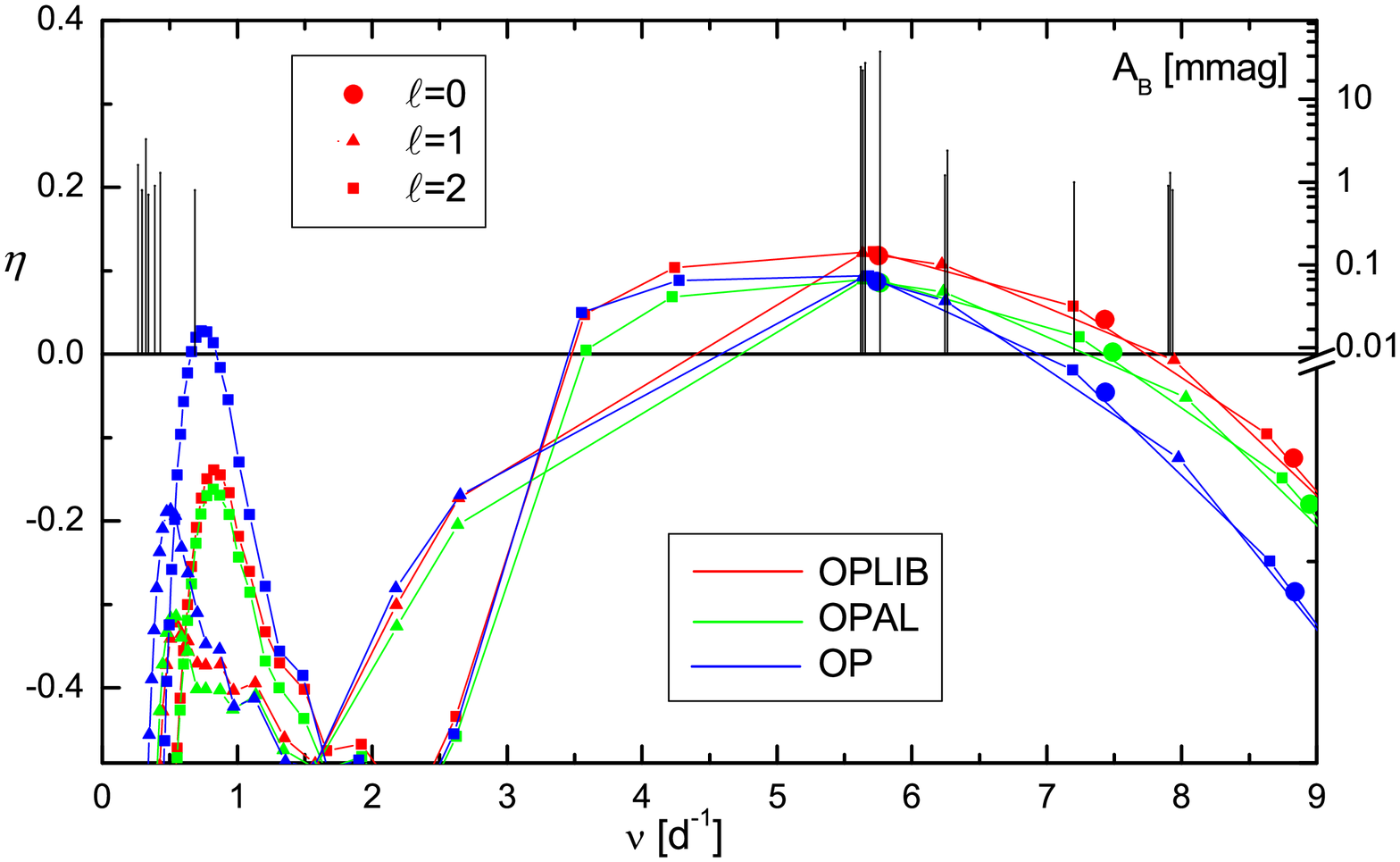}
	\caption{The normalized instability parameter, $\eta$, for representative seismic models of $\nu$ Eridani, computed with
the three opacity data: OPLIB, OPAL and OP. All models have  $M=9.5~M_\odot$, $\log T_{\rm eff}\approx 4.343$,
$\log L/L_\odot\approx 3.92$, $Z=0.015$ and the overshooting parameter $\alpha_{\rm ov}\approx 0.08$. }
	\label{fig2}
\end{figure}
\begin{figure*}
	\centering
\includegraphics[clip,width=0.48\linewidth]{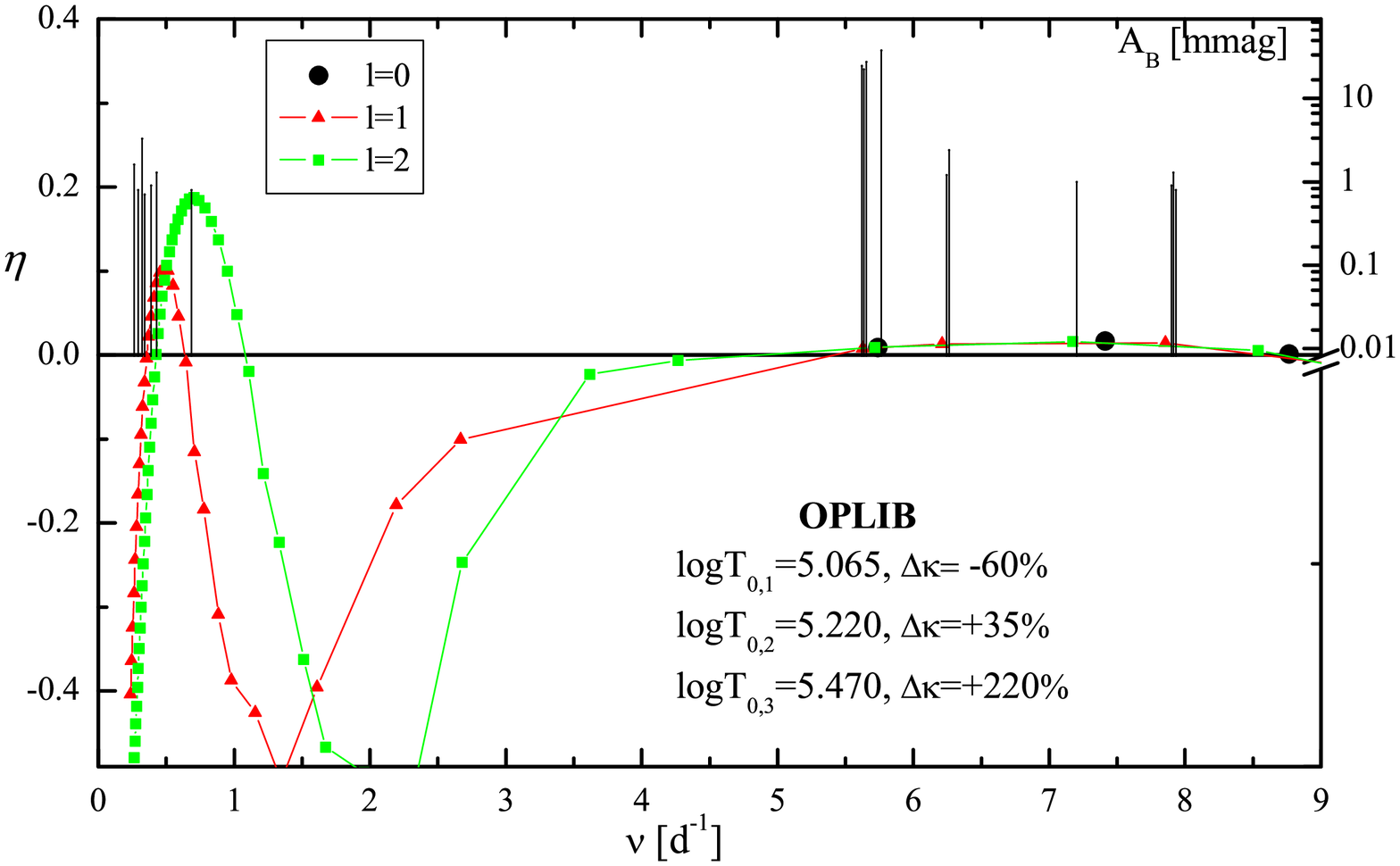}
\includegraphics[clip,width=0.48\linewidth]{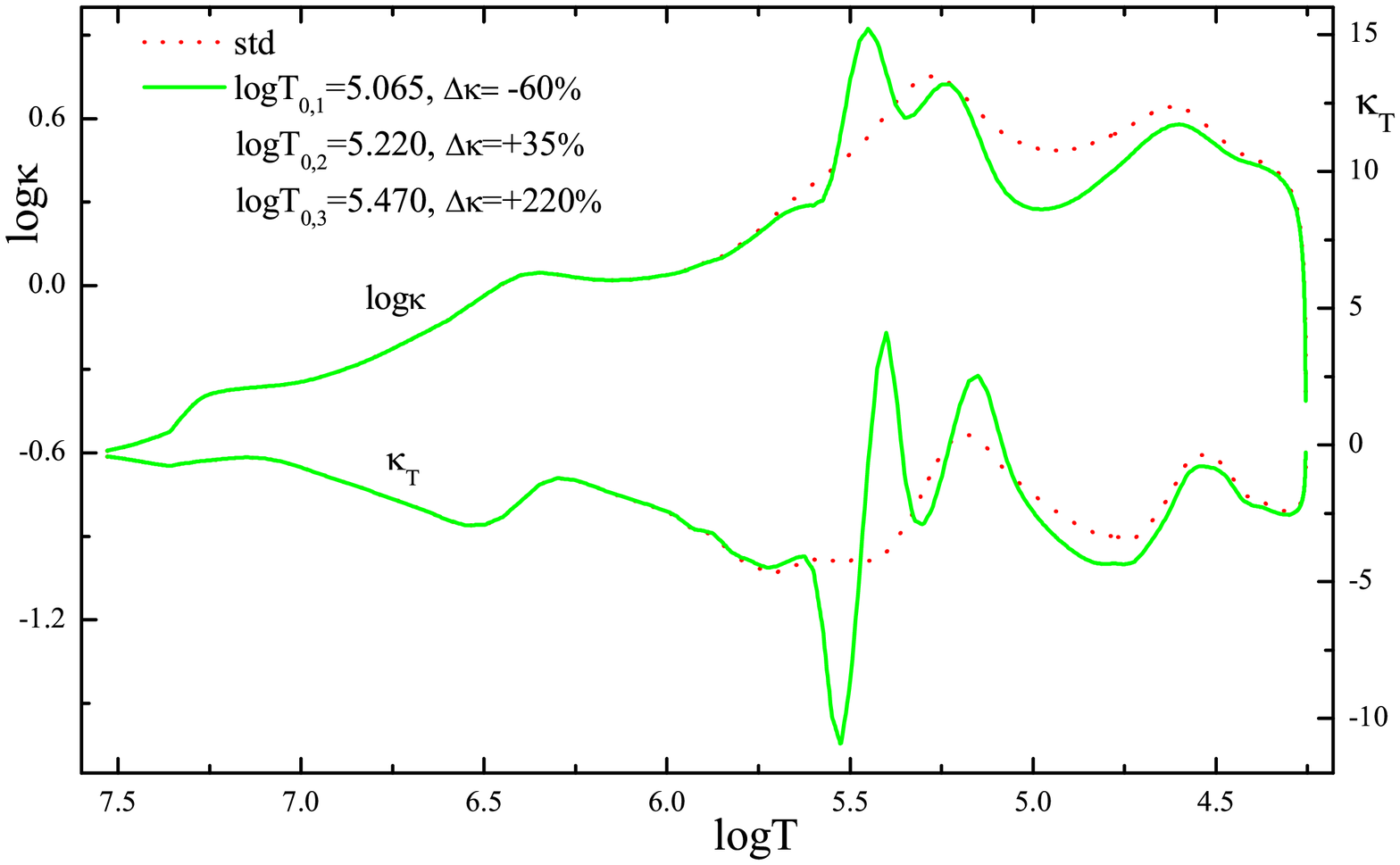}
	\caption{{\it The left panel}: the OPLIB seismic model of $\nu$ Eri, which reproduces the observed frequencies of p modes, the range of instability for
both p and g modes, and the empirical value of the nonadiabatic parameter $f$ for the dominant frequency corresponding
to the radial fundamental mode. The parameters of the model are: $M=9.0~M_\odot$, $\log T_{\rm eff}=4.3314$, $Z=0.015$, $X_0=0.7$,
$\alpha_{\rm ov}=0.163$. The modifications of opacity are given in the legend. {\it The right panel}: the corresponding modified profile of the Rosseland mean opacity (the left-hand Y-axis) and its temperature derivative (the right-hand Y-axis).}
	\label{fig3}
\end{figure*}
\begin{table*}[t]
	\centering
	\caption{The opacity modifications obtained from seismic studies of early B-type pulsators dominated by p modes. 
	The mass and effective temperature of the $\kappa-$modified models are given in the second and third column, respectively. }
	\label{table1}
	\begin{tabular*}{0.8\linewidth}{l c c c c c c c c}
	\noalign{\smallskip}\hline\hline\noalign{\smallskip}
  star & $M/M_\odot$ &  $\log (T/{\rm K})$ & $\log T_{0,1}$ & $\Delta\kappa$ & $\log T_{0,2}$ & $\Delta\kappa$ & $\log T_{0,3}$ & $\Delta\kappa$  \\
	\noalign{\smallskip}\hline\noalign{\smallskip}
	$\nu$ Eridani &  9.0 &  4.331 & 5.06 & $-60\%$ &  5.22  &  $+35\%$ & 5.46 & $+220\%$ \\
& & & & & & \\
	 12 Lacertae  & 11.2 &  4.376 & 5.06 & $-25\%$ &  5.22  &  $+50\%$ & 5.46 & $+200\%$ \\
& & & & & & \\
  $\gamma$ Pegasi & 8.1 & 4.324 & 5.06 & $-60\%$ &  5.22  &  $+50\%$ & 5.46 & $+210\%$ \\
  & & & & & & \\
$\theta$ Ophuichi & 8.4 & 4.343 & 5.06 & $+30\%$ &  5.30  &  $+65\%$ & 5.46 & $+145\%$ \\
& & & & & & \\
 $\kappa$ Scorpii & 10.4 & 4.363 & 5.06 & $+30\%$ &  5.22  &  $+30\%$ & 5.46 & $+100\%$ \\
 & & & & & & \\
$\alpha$ Lupi & 12.0  & 4.351 & -- & -- &  --  &  -- & 5.46 & $+100\%$ \\
	\noalign{\smallskip}\hline
	\end{tabular*}
\end{table*}
It turned out that fitting the parameter $f$ significantly reduces the number of seismic models and, in fact, it is quite hard
to find the models that simultaneously reproduce the observed frequencies, the range of instability for both p and g modes
and the empirical value of the nonadiabatic parameter $f$. This is a good news because it means that the parameter $f$ has
a great diagnostic power for constraining stellar opacities in B-type pulsators.
Such complex OPLIB seismic model is shown in Fig.\,3, where the instability parameter, $\eta$,
is plotted as a function of the mode frequency on the left panel and the corresponding mean opacity profile is compared 
with the standard one on the right panel. The model has a huge increase (by about 220\%) of opacity at the temperature $\log T=5.46$ 
where nickel has its maximum contribution to the mean opacities. This modification was enforced by the requirement of instability
of low-frequency modes. The opacity increase at this temperature was also suggested by \cite{Salmon2012}
who studied B-type pulsators in the Magellanic Clouds.
Besides, adding the opacity at $\log T=5.22$ was necessary to keep the instability of p modes in the observed frequency range.
Finally, to fit the parameter $f$ we had to reduce the opacity in more shallower layers of the envelope, i.e., at $\log T =5.06$. 
The last modification was quite surprising because just for the temperature $\log T=5.06$, \cite{Cugier2014} identified 
an additional local maximum from Kurucz model atmospheres.  Meanwhile, we do not have a good explanation for this result.
The instability of the lowest frequency modes in this model can be explained by including the rotational splitting. 

All our results obtained for $\nu$ Eri were published in \cite{JDD2017} and here we only briefly summarized
the main steps and conclusions. We refer the reader to this paper for more details.

In a similar way we studied other early B-type stars: 12 Lac, $\gamma$ Peg,  $\theta$ Oph, $\kappa$ Sco and  $\alpha$ Lup.
In the case of the last two stars we could not use the empirical values of $f$ because there were no proper data to determine it.
The summary of the opacity modifications derived from these seismic studies is given in Table\,1. 
More details on them can be found in \cite{Walczak2017} and \cite{JDD2018} or will be published soon (Walczak et al., in preparation).

\subsection{G-mode dominated pulsators}

In this subsection, we will present two pulsators with the oscillation spectrum dominated by high-order g modes,
in the sense that their amplitudes are much larger than the amplitudes of p modes.

Here, we used MESA evolutionary code \citep{Paxton2011, Paxton2013} and Warsaw nonadiabatic code for linear pulsations 
\citep{Dziembowski1977} in its version including the effects of rotation on pulsations \citep{Dziembowski2007}.
Because the g-mode frequencies can be of the order of the rotation frequency, including the effects of rotation on pulsations is unavoidable. 
This has been done in the framework of the traditional  approximation \citep[e.g.,][]{Bildsten1996, Lee1997, Townsend2003}
which takes into account the effects of the Coriolis force.

KIC3240411 is the hottest  known SPB/$\beta$ Cep star with the asymptotic g-mode period spacing. 
Its effective temperature is about $T_{\rm eff}= 21~000$ K and the spectral type B2V.
The star was observed by {\it Kepler} and classified as the SPB/$\beta$ Cep hybrid pulsator by Balona et al. (2011, 2015) 
who found in total more than 100 frequency peaks  with the most prominent ones below 2 d$^{-1}$
and many very low-amplitude frequency peaks up to 17 d$^{-1}$.  We performed an independent frequency analysis and found
a sequence of 22 frequency peaks which are (quasi)equally spaced in period \citep{Szewczuk2018}.
The mean period spacing of this sequence is about $\overline{\Delta P}=0.03$\,d.  In Fig.\,4, we plotted the amplitude spectrum 
and the values of $\Delta P$ as a function of period in the top and bottom panel, respectively.
The period spacing pattern predicted by the asymptotic theory enables simultaneous mode identification and seismic modelling.

We constructed an extensive grid of seismic models changing mass, metallicity, hydrogen abundance, rotation 
and overshooting parameter. All modes with $\ell\le 3$ were considered. 
It turned out that observations are best reproduced if the period spacing is associated with consecutive dipole axisymmetric modes ($\ell=1,~m=0$).
In the next step, we checked instability of the modes in the observed frequency range. Because KIC3240411 is the early B-type star, 
it shares the same problem as pulsators described in the previous subsection, i.e., all standard opacity models are pulsationally stable 
in the low-frequency range.  Therefore, to account for mode instability, we modified the mean opacity profile. 
In Fig.\,5, we show a comparison of the observed frequencies with the theoretical ones for five seismic models.
The parameters of the models are given in Table\,2 and the opacity modifications in Table\,3.
The model $\#22$ gives the best fit of the observed period spacing.

\begin{table*}
	\centering
	\caption{The parameters of the best seismic models of KIC3240411 reproducing the sequence of the period spacing.
	The subsequent columns contain: the mass, the current value of rotation, the rotation frequency, the initial abundance of hydrogen, 
	metallicity, overshooting parameter, effective temperature, luminosity, surface gravity, central hydrogen abundance and the maximum value 
	of the parameter $\eta$ for the modes associated with the observed sequence. In the last column, there are model numbers.}
	\label{tab_best_mod_OPLIB}
	\begin{tabular}{cccccccccccccccc}
\hline	
 $M/M_\odot$     &  $V_{\rm rot}$  & $\nu_{\rm rot}$ & $X$ & $Z$ & $f_{\rm ov}$ &  $\log T_{\rm eff}$ & $\log L/L_\odot$ & $\log g$ & $X_c$ & $\eta_{\rm max}$ & opac  & mod. \\
    & [km\,s$^{-1}$] &  [d$^{-1}$] & &  &  &   &    &  &  &  &  & \\
\hline
 6.25 & 117.35 & 0.82 & 0.670 & 0.006 & 0.02  & 4.3486 & 3.255 & 4.327 & 0.612  &   -0.39   &   std.  & \#2    \\
 7.95 & 171.65 & 0.95 & 0.738 & 0.013 & 0.02  & 4.3402 & 3.421 & 4.231 & 0.676  &   -0.29 &     std. &  \#4 \\
\hline
7.80 & 164.35 & 0.92 & 0.738 & 0.013 & 0.02  & 4.3350 & 3.391 & 4.233 & 0.681  &   -0.11 &  O4 & \#15\\
6.25 & 124.05 & 0.80 & 0.738 & 0.013 & 0.02  & 4.2832 & 3.062 & 4.258 & 0.686  &   0.23 &  O5 & \#16\\
6.35 & 124.00 & 0.80 & 0.740 & 0.011 & 0.015 & 4.2942 & 3.105 & 4.266 & 0.676  &  0.18  &  O8 & \#22\\
\hline
\end{tabular}
\end{table*}

\begin{table}
	\centering
	\caption{The opacity modification coding for seismic models of KIC3240411.}
	\label{tab_opac_modifications}
	\begin{tabular}{@{}cc@{}cc@{}cc@{}c@{}c@{}c@{}}
\hline	
opac &  $\log T_{0,1}$ & $\Delta\kappa$ & $\log T_{0,2}$ & $\Delta\kappa$ & $\log T_{0,3}$ & $\Delta\kappa$  \\
	\hline
O4  & 5.06 &  $+100$\%  & 5.30 &  $+100$\% & 5.46 &  $+100$\%   \\
O5  & 5.46 &  $+200$\%  &    --     &     --              &   --      &  --             \\
O8  & 5.06 &   $-50$\%    & 5.22 &  $+50\%$   & 5.46 & $+200$\%   \\
\hline
\end{tabular}
\end{table}

Details of the frequency analysis and seismic studies of KIC3240411 can be found in \cite{Szewczuk2018}.
Here, in figures and tables, we left a notation from this paper.

\begin{figure}
	\centering
\includegraphics[clip,width=0.75\linewidth, angle=-90]{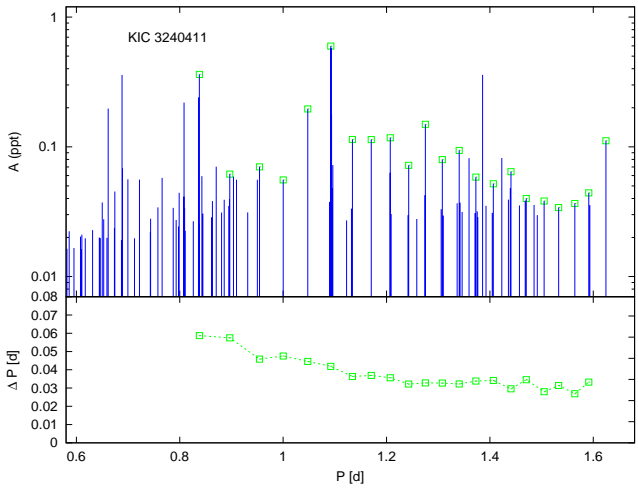}
	\caption{A part of the oscillation spectrum of KIC3240411 where quasi-equidistant structures in period were found ({\it the top panel}).
	The peaks belonging to this sequence are marked with squares. The corresponding period spacing
	 as a function of the period is shown in {\it the bottom panel}.}
	\label{fig4}
\end{figure}
\begin{figure}
	\centering
\includegraphics[clip,width=\linewidth]{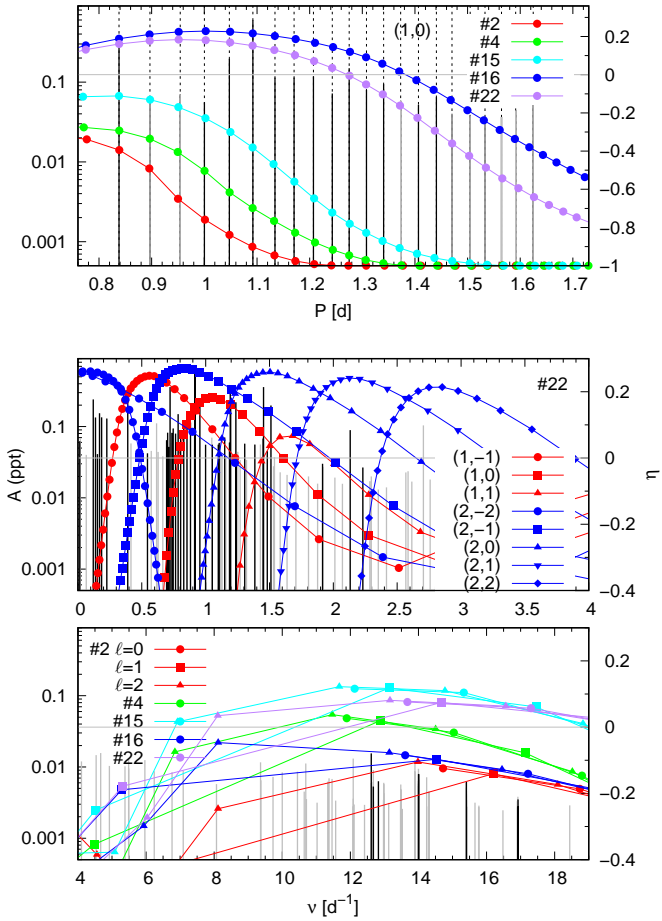}
	\caption{{\it The top panel}: the amplitude spectrum of frequencies forming the period-spaced sequence in comparison with 
	the theoretical frequencies of axisymmetric dipole modes ($\ell=1,~m=0$) for selected best seismic models of KIC3240411. 
	The values of the mode instability parameter, $\eta$, are on the right-hand Y-axis. The observed frequencies which are 
	definitely independent are marked by the black vertical lines and possible combinations as the grey lines.
        {\it The middle panel}: All observed frequencies in the g-mode range compared with dipole and quadrupole modes of the model \#22.
        {\it The bottom panel}: The observed frequencies and their amplitudes in the p-mode range confronted with the theoretical values 
        from models considered in the top panel.}
	\label{fig5}
\end{figure}

The second example is KIC11971405,  the rapidly rotating star of B5 IV-Ve spectral type. 
The basic stellar parameters, as determined by \cite{Papics2017} from spectroscopic observations, 
are: the effective temperature, $\log T_{\rm eff} = 4.179(6)$, surface gravity, $\log g = 3.94(6)$
and the projected rotation velocity, $V_{\rm rot} \sin i = 242(14)$ km$\cdot$s$^{-1}$.

In the {\it Kepler} photometric data \cite{Papics2017} found three period--spaced sequences around 
$P=0.53,~0.45$ and 0.27 days with the mean period spacing of about $\overline{\Delta P}=0.0025$, 0.0032 and 0.001\,d, respectively.
From our seismic modelling we got for the sequence around  $P=0.53$ d the mode identification $(\ell=1,~m=+1)$,
for the sequence around $P=0.27$ d -- the identification $(\ell=2,~m=+2)$.  We called these sequences $s1$ and $s2$, respectively.
The third sequence around $P=0.45$\,d could not be reproduced by any mode and we considered it accidental.
In the top panel of Fig.\,6, we depicted  the observed amplitude spectrum as a function of period around the two period-spaced sequences.
The corresponding values of $\Delta P$ are given in the bottom panel. 
\begin{figure}
	\centering
\includegraphics[clip,width=0.75\linewidth, angle=-90]{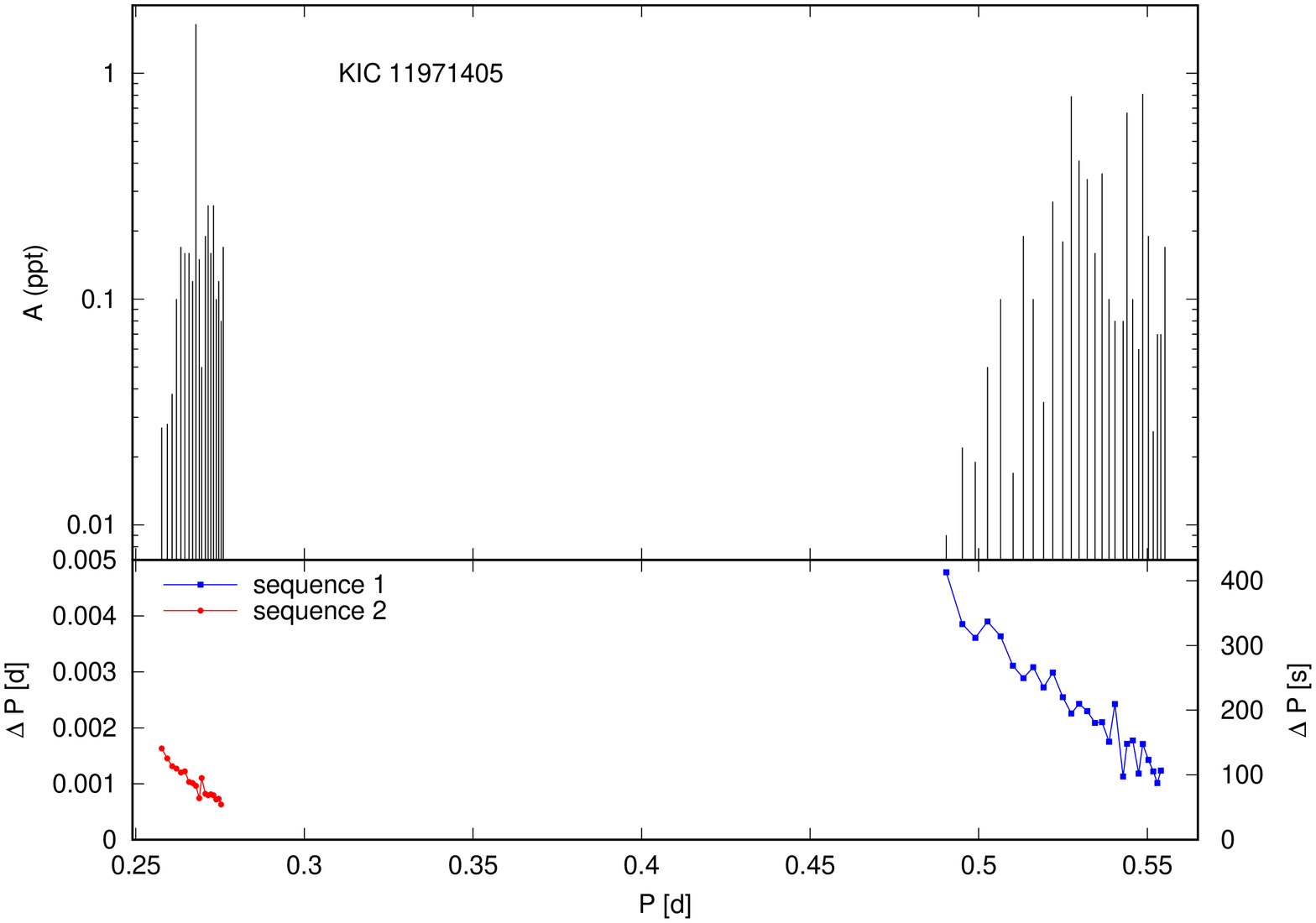}
	\caption{A part of the oscillation spectrum of KIC11971405 where quasi-equidistant structures in period were found ({\it the top panel})
	and the period spacing, $\Delta P$, of the identified sequences as a function of the period ({\it the bottom panel}).}
	\label{fig6}
\end{figure}
\begin{figure}
	\centering
\includegraphics[clip,width=0.75\linewidth, angle=-90]{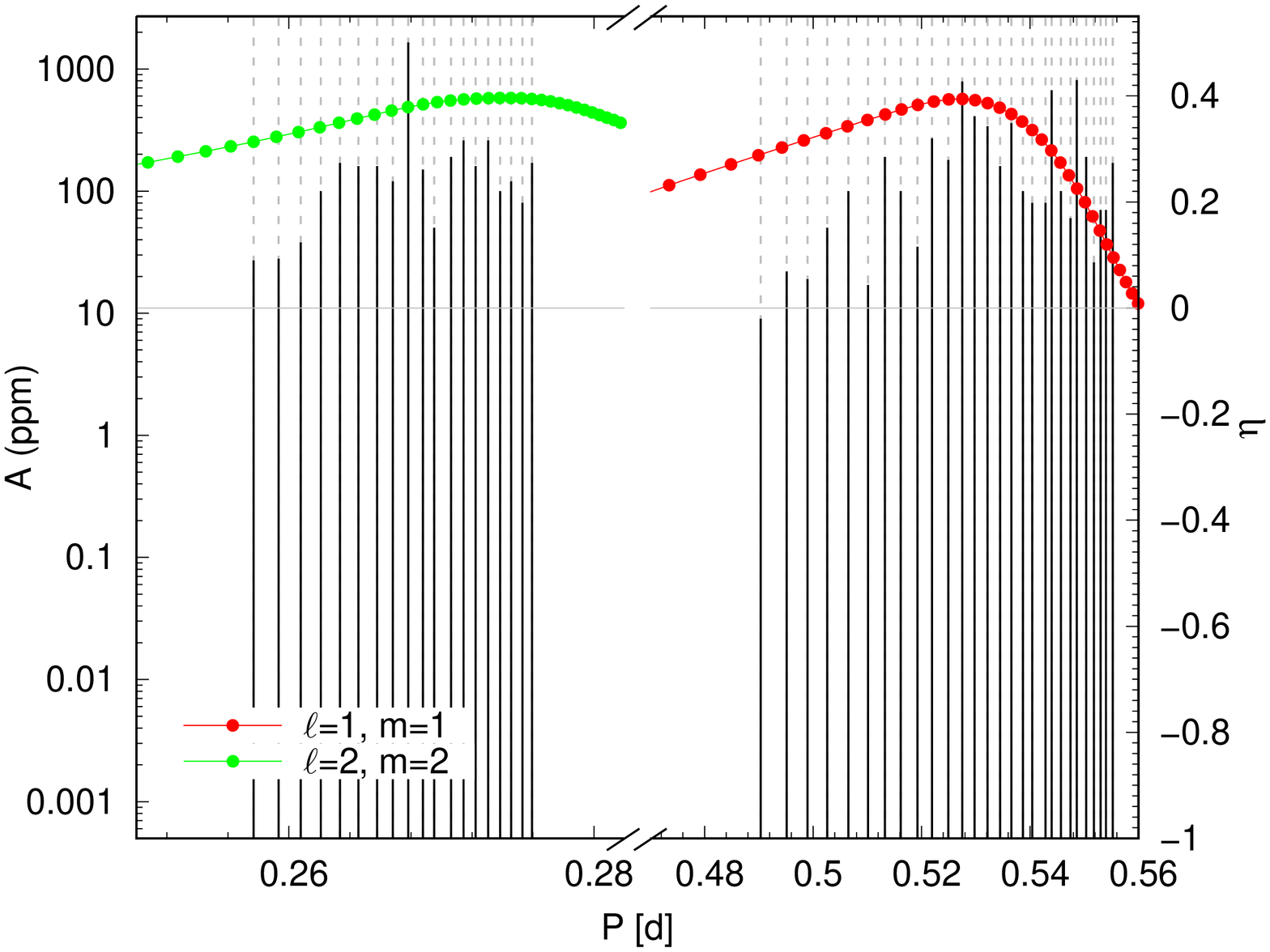}
	\caption{The oscillation spectrum of KIC11971405 in the range of the two sequences which are (quasi)equally spaced in period. 
	There is shown also the $\kappa$-modified OPLIB seismic model which reproduces simultaneous the two sequences which can be 
	accounted for only by prograde sectoral modes with $\ell=1$ and $\ell=2$. }
	\label{fig7}
\end{figure}

As far as we are aware, this is the first B-type stars exhibiting the two period-spacing structures. It gives prospects for 
more stringent constraints from seismic analysis.  Using the standard opacity data, we encountered a problem with mode excitation
for modes with period greater that about $P=0.51$\,d, i.e., the lowest frequency peaks, which are part of the sequence $s1$. 
To get the instability in the whole range of the sequence $s1$, we modified the mean opacity. As an example, in Fig.\,7  
we show a comparison of the observed oscillation spectrum in the range of the sequences $s1$ and $s2$
with the theoretical model reproducing the period spacing of both sequences and accounting for their instability.  The instability of these modes 
has been achieved by an opacity increase at temperatures $\log T=5.3$ and $\log T=5.46$ by 50\% and 200\%, respectively.
Moreover, a decrease of opacity at logT=5.06 by 50\% was needed.

More details on seismic analysis of KIC11971405 will be presented by Szewczuk et al. (in preparation).

\section{Constraints on convective overshooting}

The overshooting parameter gives the information on the amount of matter mixed above convective regions.
It is a very important parameter in evolutionary modelling but unfortunately it cannot be derived from first principles
and is described by the free parameter.
\begin{table*}[t]
	\centering
	\caption{The values of the overshooting parameter obtained for the analyzed early B-type pulsators.}
	\label{table4}
	\begin{tabular*}{0.5\linewidth}{l c c c c c }
	\noalign{\smallskip}\hline\hline\noalign{\smallskip}
	  $\nu$ Eri & 12 Lac & $\gamma$ Peg & $\theta$ Oph & $\kappa$ Sco  \\
	\noalign{\smallskip}\hline\noalign{\smallskip}
	 $(0-0.16)$  & $(0.2-0.5)$ & $(0.2-0.3)$ &  $\simeq 0.3$  &  $\simeq 0.2$  \\
	\noalign{\smallskip}\hline
	\end{tabular*}
\end{table*}

Seismic studies of pulsating stars has the greatest potential to obtain reliable constraints on the overshooting parameter. 
Some additional information comes also from studies of double-lined eclipsing binaries \citep[e.g.,][]{Higl2017}
but they cannot be treated as precise estimates. The first stringent value of overshooting from the convective core was obtained 
for the $\beta$ Cep star HD129929 \citep{Aerts2003}. Thereafter, there was a flood of publications on this subject and there is no space here
to mention all of them. The typical value of overshooting from seismic studies is below $0.3H_p$, where $H_p$ is the pressure scale height.

In Table\,4, we summarize the values of the overshooting parameter, $\alpha_{\rm ov}$, obtained from our seismic studies
of early B-type stars dominated by p modes. In the Warsaw-New Jersey code overshooting is described by a two-parameter prescription 
that allows for non-zero gradient of the hydrogen abundance inside the partly mixed region above the convective core 
\citep{Dziembowski2008}. As one can see in each case we got $\alpha_{\rm ov}\lesssim 0.3$.

Constraints on the overshooting parameter should be stronger for g-mode dominated pulsators, because these modes
have higher amplitudes in deeper interior and are more sensitive to conditions near the core. However,   
because in these cases taking into account the effects of rotation is inevitable, the allowed range of the overshooting parameter
is usually wider.  

In Fig.\,8, we plotted the values of the discriminant $\chi^2$, giving the goodness of the fit between the theoretical and observed frequencies,
as a function of the overshooting parameter, $f_{\rm ov}$, as denoted in MESA code, for seismic models of KIC3240411. 
From this dependence one can conclude that the overshooting parameter from the convective core of KIC3240411 should not be larger
than $f_{\rm ov}=0.025$.
In MESA code overshooting is treated as a diffusive process and it is described by the exponential formula \citep{Paxton2011}.
The relation  between the MESA and Warsaw overshooting parameters is $\alpha_{\rm ov}\approx10f_{\rm ov}$.

\begin{figure}
	\centering
\includegraphics[clip,width=\linewidth, angle=-90]{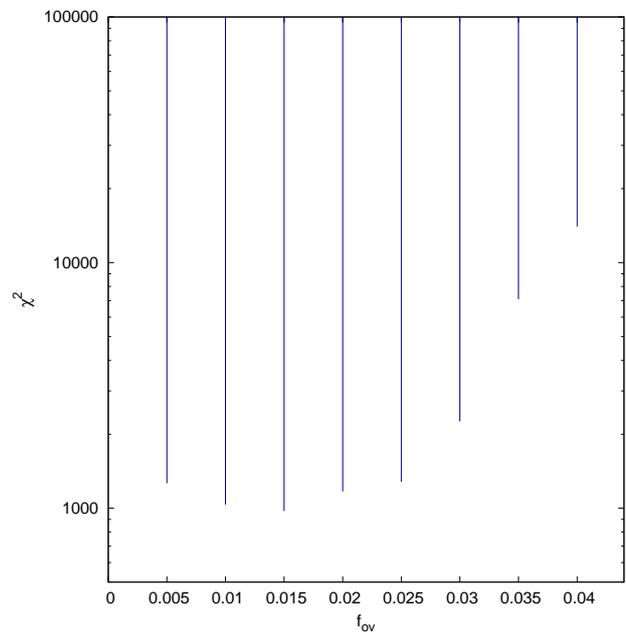}
	\caption{The values $\chi^2$ as a function of the overshooting parameter, $f_{\rm ov}$, for seismic models of KIC3240411.}
	\label{fig8}
\end{figure}

The case of KIC11971405 was more complicated because of a very high rotation, $V_{\rm rot}\gtrsim 230$ km\,s$^{-1}$.
If only one sequence is considered, then there is no good discrimination of the parameter $f_{\rm ov}$,
regardless of whether the rotation is rigid or differential. This is illustrated in the top and middle panels of Fig.\,9.  
Changing the parameters of rotation-induced mixing, e.g., enhancing the Eddington--Sweet circulation, did not improve the result. 
Always a weak sensitivity to convective overshooting was obtained. 
Definitely better constraints on $f_{\rm ov}$ were derived if the two sequences, $s1$ and $s2$, were fitted simultaneously,
as shown in the bottom panel of Fig.\,9. However, there is still no single, clear minimum of $\chi^2(f_{\rm ov})$.
This can be caused by a very fast rotation that blurs the effect of overshooting mixing.
\begin{figure}
	\centering
\includegraphics[clip,width=1.35\linewidth, angle=-90]{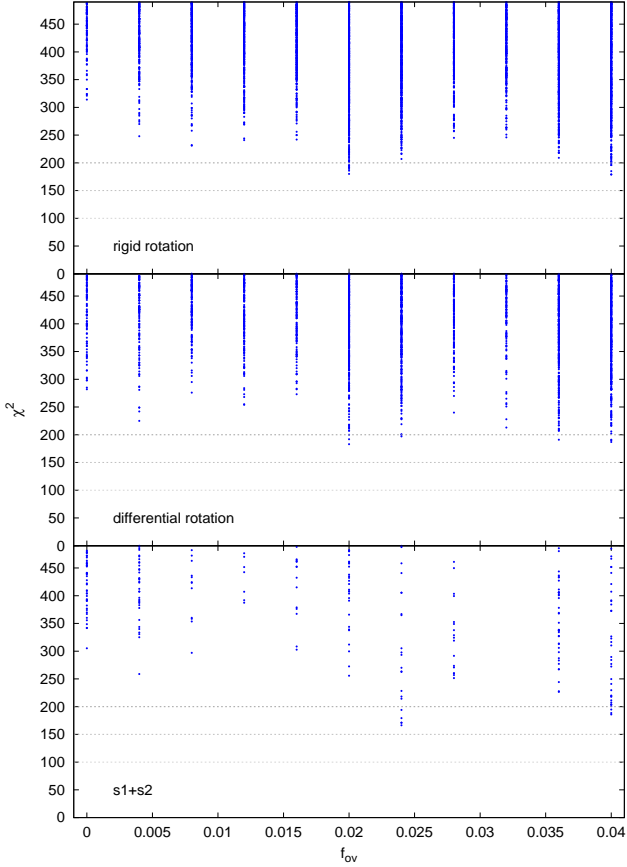}
	\caption{The values $\chi^2$ as a function of the overshooting parameter, $f_{\rm ov}$, for seismic models of KIC11971405. The models in the top and middle panels 
	reproduce only the sequences $s1$ and were computed at the assumptions of the rigid (top) and differential (middle) rotation. In the bottom panel, 
	the values of $\chi^2$ for seismic models reproducing both sequences, $s1$ and $s2$, are plotted. These models rotate differentially.}
	\label{fig9}
\end{figure}

\section{Constraints on rotation}

Rotation is another main agent that controls stellar evolution and co--determines the ultimate fate of a star.
It is not only about the value of surface rotation but also about the exact rotational profile inside the star.

An interesting result was obtained for $\theta$ Oph A from a requirement of mode instability in the whole g-mode frequency range (Walczak et al.,
in preparation).
Employing the traditional approximation, we obtained that, besides increasing opacities, the internal rotation has to be at least two times
larger than the surface value resulting from the rotational splitting of p modes.  This indicates that the star can rotate differentially. 
This result is presented in Fig.\,10,  where the theoretical frequencies and their instabilities are compared with the observed frequencies.
The top panel shows the model computed with the rotation $V_{\rm{rot}}=28$ kms$^{-1}$ whereas the bottom panel with $V_{\rm{rot}}=65$ km\,s$^{-1}$.
\begin{figure}
	\centering
\includegraphics[clip,width=\linewidth]{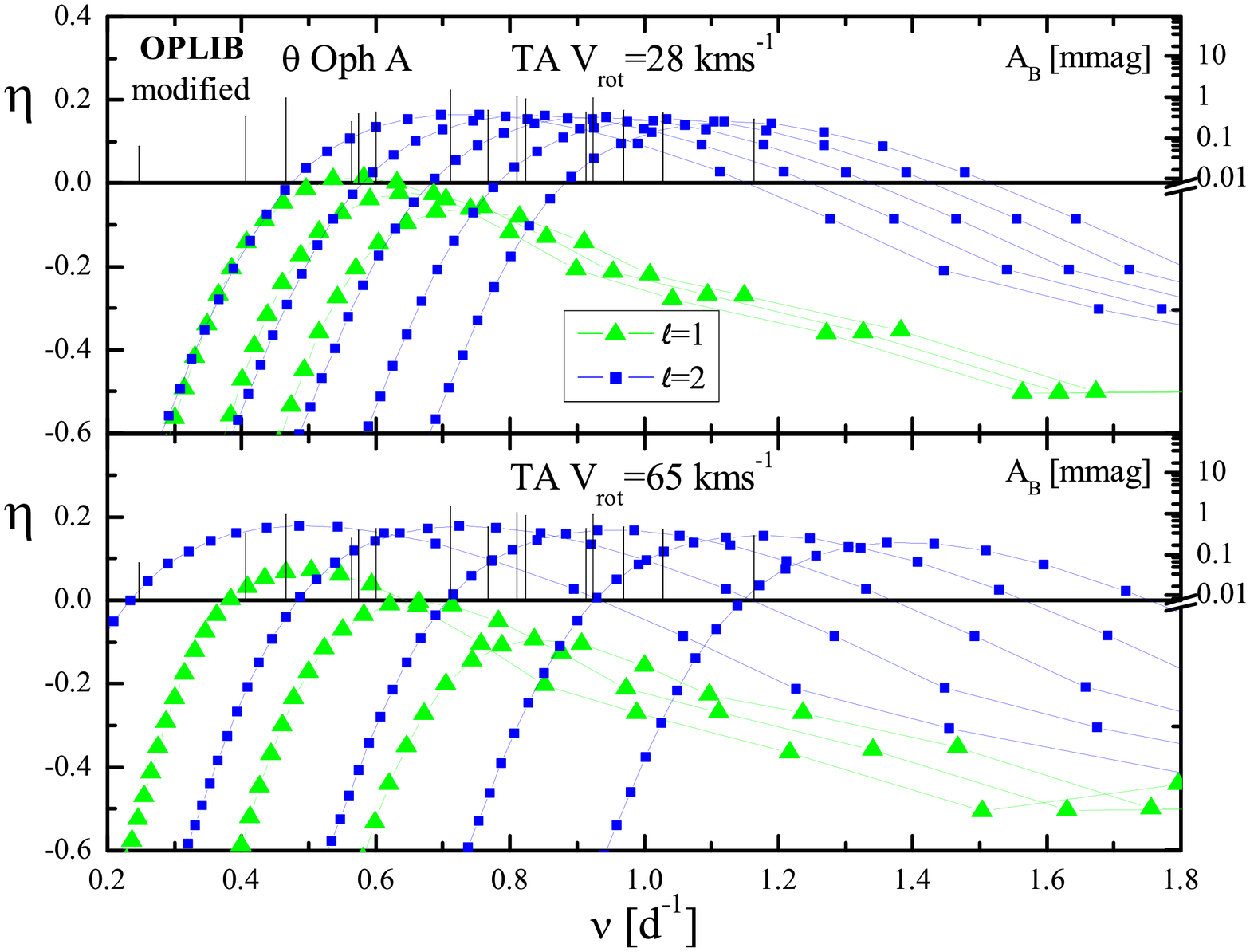}
\caption{Comparison of the oscillation frequency  of $\theta$ Oph A in the g-mode range with the instability range of pulsational modes $\ell=1$ and 2 
in the $\kappa-$modified seismic model. 
Two values of rotational velocity were assumed: $V_{\rm{rot}}=28$ kms$^{-1}$ (the top panel) and $V_{\rm{rot}}=65$ kms$^{-1}$ (the bottom panel).}
	\label{fig10}
\end{figure}

For g-mode pulsator KIC3240411, we obtained the rotation velocity in the range 160--180 km\,s$^{-1}$. The result is shown 
in Fig.\,11 where the values of $\chi^2$ of seismic models are plotted as a function of $V_{\rm rot}$.
This range of rotation refers rather to the interior of the star, because it was derived from high-order g modes which have 
the largest amplitudes in the deeper interior.
Given the value of the surface projected rotational velocity, $V_{\rm rot}\sin i = 43(5)$ km\,s$^{-1}$, it would mean that the internal rotation
of KIC3240411 is about 4 times faster, if the inclination angle is not far from $i=90^\circ$.
\begin{figure}
	\centering
\includegraphics[clip,width=\linewidth, angle=-90]{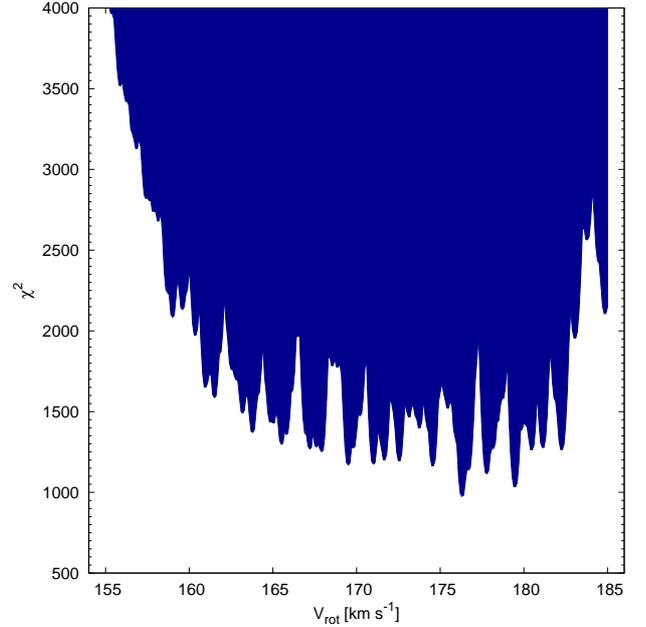}
	\caption{The values $\chi^2$ as a function of the rotational velocity, $V_{\rm rot}$, for seismic models of KIC3240411.}
	\label{fig11}
\end{figure}

In the case of KIC11971405 constraints on rotation are even worse than on convective overshooting. 
In Fig.\,12, we show the values of $\chi^2$ as a function of  $V_{\rm rot}$, for seismic models of KIC11971405 fitting 
the period spacing $s1$. We considered three types of rotation: rigid, differential and with a three-fold increase of mixing
from Eddington--Sweet (meridional) circulation.  As one can see, no case gives discrimination of the rotational velocity.
The reason could be too fast rotation at which the traditional approximation may be no longer valid. 

\begin{figure}
	\centering
\includegraphics[clip,width=1.35\linewidth, angle=-90]{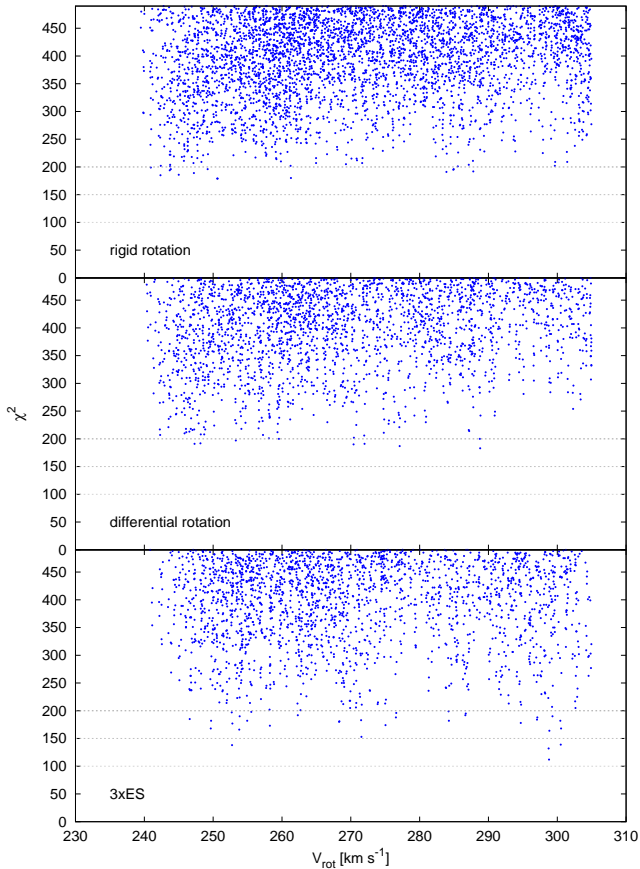}
	\caption{The values of $\chi^2$ plotted as a function of rotation, $V_{\rm rot}$, for seismic models of KIC11971405 fitting the period spacing $s1$.
	Three types of rotation were considered: rigid (the top panel), differential (the middle panel) and with a three-fold 
	increase of meridional circulation (ES stands for Eddington--Sweet).}
	\label{fig12}
\end{figure}

\section{Summary and further prospects}

The main goal of asteroseismic studies is to obtain constraints on physical conditions inside the stars.
To this aim, the values of parameters of the model and theory are estimated. 
In this paper, we presented the outline of our results obtained from seismic studies of B-type pulsators.
In the following sections, we described what we have learned about stellar opacities, overshooting from the convective core
and rotation from pulsations of B-type main sequence stars. 

We learned that stellar opacities in the vicinity of metal bump ($\log T=5.3$) still need further modifications in order to account 
for instability of all observed modes. In particular, to excite high-order g modes  in early B-type stars one requires a huge increase (about 200\%)
of opacity at the depth corresponding to the temperature $\log T=5.46$.  At this temperature nickel has its maximum contribution to the $Z-$bump.
Additionally, some opacity increase (about 50\%) is needed around $\log T=5.2-5.3$.
What is the reason of such a huge increase of stellar opacities? The significant increase of the opacity coefficient has been obtained recently
by \cite{HuiBonHoa2018} through diffusion which caused the accumulation of iron and nickel in the driving zone. 
Therefore, diffusion can be a partial explanation of g-modes instability problem. Definitely, further and more detailed studies are necessary
in that direction.

To control the number of opacity modifications the usage of the nonadiabatic parameter $f$ is crucial. This parameter is 
a very powerful seismic probe of subphotospheric layers where pulsations of B-type stars are driven. The important result is that it is very
hard to find the model that simultaneously fits the observed frequencies, accounts for the instability in the whole frequency range and
reproduces the empirical values of $f$.   
Another important modification, we obtained in the case of a few stars, was a reduction of opacity at the temperature $\log T=5.06$.
This result is quite perplexing because at this temperature \cite{Cugier2014} uncovered the new opacity bump in Kurucz model atmospheres.

Seismic constraints on opacities are of much greater importance because they provide an independent test of these microphysics data.

Another puzzle in evolutionary modelling is overshooting from convective zones. 
Our seismic studies provided constraints on overshooting from the convective core.
The efficiency of overshooting is described by the free parameter, denoted as $\alpha_{\rm ov}$ or $f_{\rm ov}$ depending on the code.
Up to now asteroseismology is the most stringent method to derive its values. 
The general result from our seismic analysis of the presented pulsators is that the value of overshooting should not be larger than about $0.3H_p$.
In the case of the very fast rotating star, KIC11971405, all seismic models are very poorly sensitive to overshooting from the convective core.
It could be that at very high rotation the effects of convective overshooting is completely blurred by rotation-induced mixing.

The last results concern rotation. For massive star evolution rotation is as important as mass and metallicity.
Moreover, the knowledge of the rotational profile at various evolutionary stages could give an indication
of how the angular momentum transport in stars could occur.  An excellent review on the angular momentum transport in stellar interiors
and constraining it from asteroseimic studies has been published recently by \cite{Aerts2018}.

From instability analysis of g modes in the early B-type star $\theta$ Oph A, we obtained that its core should rotate at least twice as fast as the envelope.
In the case of the star KIC3240411, exhibiting the period-spaced structure in the oscillation spectrum, the core can rotate up to four times faster. 
On the other hand, no constraint on rotation was obtained from seismic studies of KIC11971405. 
The reason could be again a very fast rotation 
of the star ($V_{\rm rot}>230$ km\,s$^{-1}$) at which the traditional approximation ceases to apply. 

Despite some limitations and simplifications in our modelling, we believe that the presented 
results can contribute to a better understanding of the internal structure of the stars and further improvement of evolutionary codes.
We intend to continue these seismic studies and extend them to lower-mass pulsators, e.g., $\delta$ Scuti stars.

\section*{Acknowledgments}
The work was financially supported by the Polish NCN grant 2015/17/B/ST9/02082 and DEC-2013/08/S/ST9/00583.

\bibliographystyle{phostproc}
\bibliography{JDDbiblio.bib}

\begin{thebibliography}{41}
\providecommand{\natexlab}[1]{#1}

\bibitem[\protect\astroncite{{Aerts} \emph{et~al.}}{2018}]{Aerts2018}
{Aerts}, C., {Mathis}, S., \& {Rogers}, T. 2018, ArXiv e-prints.

\bibitem[\protect\astroncite{{Aerts} \emph{et~al.}}{2003}]{Aerts2003}
{Aerts}, C., {Thoul}, A., {Daszy\'nska}, J., \& {et al.} 2003, Science, 300,
  1926.

\bibitem[\protect\astroncite{{Bailey} \emph{et~al.}}{2015}]{Bailey2015}
{Bailey}, J.~E., {Nagayama}, T., {Loisel}, G.~P., \& {et al.} 2015, Nature,
  517, 56.

\bibitem[\protect\astroncite{{Balona} \emph{et~al.}}{2015}]{Balona2015}
{Balona}, L., {Baran}, A., {Daszy\'nska-Daszkiewicz}, J., \& {De Cat}, P. 2015,
  MNRAS, 451, 1445.

\bibitem[\protect\astroncite{{Balona} \emph{et~al.}}{2011}]{Balona2011}
{Balona}, L., {Pigulski}, A., {De Cat}, P., \& {et al.} 2011, MNRAS, 413, 2403.

\bibitem[\protect\astroncite{{Bildsten} \emph{et~al.}}{1996}]{Bildsten1996}
{Bildsten}, L., {Ushomirsky}, G., \& {Cutler}, C. 1996, ApJ, 460, 827.

\bibitem[\protect\astroncite{{Blancard} \emph{et~al.}}{2016}]{Blancard2016}
{Blancard}, C., {Colgan}, J., {Cosse}, P., \& {et al.} 2016, Phys.Rev.Letters,
  117, 249501.

\bibitem[\protect\astroncite{{Colgan} \emph{et~al.}}{2015}]{Colgan2015}
{Colgan}, J., {Kilcrease}, D.~P., {Magee}, N.~H., \& {et al.} 2015, High Energy
  Density Physics, 14, 33.

\bibitem[\protect\astroncite{{Colgan} \emph{et~al.}}{2016}]{Colgan2016}
{Colgan}, J., {Kilcrease}, D.~P., {Magee}, N.~H., \& {et al.} 2016, ApJ, 817,
  116.

\bibitem[\protect\astroncite{{Cugier}}{2014}]{Cugier2014}
{Cugier}, H. 2014, A\&A, 565, A76.

\bibitem[\protect\astroncite{{Daszy{\'n}ska-Daszkiewicz}
  \emph{et~al.}}{2003}]{JDD2003}
{Daszy{\'n}ska-Daszkiewicz}, J., {Dziembowski}, W.~A., \& {Pamyatnykh}, A.~A.
  2003, A\&A, 407, 999.

\bibitem[\protect\astroncite{{Daszy{\'n}ska-Daszkiewicz}
  \emph{et~al.}}{2005}]{JDD2005}
{Daszy{\'n}ska-Daszkiewicz}, J., {Dziembowski}, W.~A., \& {Pamyatnykh}, A.~A.
  2005, A\&A, 441, 641.

\bibitem[\protect\astroncite{{Daszy{\'n}ska-Daszkiewicz}
  \emph{et~al.}}{2017}]{JDD2017}
{Daszy{\'n}ska-Daszkiewicz}, J., {Pamyatnykh}, A.~A., {Walczak}, P., \& {et
  al.} 2017, MNRAS, 466, 2284.

\bibitem[\protect\astroncite{{Daszy{\'n}ska-Daszkiewicz}
  \emph{et~al.}}{2018}]{JDD2018}
{Daszy{\'n}ska-Daszkiewicz}, J., {Walczak}, P., {Pamyatnykh}, A.~A., \& {et
  al.} 2018, In \emph{Proceedings of the Polish Astronomical Society},
  \emph{PTA Proceedings}, vol.~8, p.~64.

\bibitem[\protect\astroncite{{Degroote} \emph{et~al.}}{2009}]{Degroote2009}
{Degroote}, P., {Aerts}, C., {Ollivier}, M., \& {et al.} 2009, A\&A, 506, 471.

\bibitem[\protect\astroncite{{Dziembowski}}{1977}]{Dziembowski1977}
{Dziembowski}, W.~A. 1977, Acta Astron., 27, 95.

\bibitem[\protect\astroncite{{Dziembowski}
  \emph{et~al.}}{2007}]{Dziembowski2007}
{Dziembowski}, W.~A., {Daszy{\'n}ska-Daszkiewicz}, J., \& {Pamyatnykh}, A.~A.
  2007, MNRAS, 374, 248.

\bibitem[\protect\astroncite{{Dziembowski}
  \emph{et~al.}}{1993}]{Dziembowski1993}
{Dziembowski}, W.~A., {Moskalik}, P., \& {Pamyatnykh}, A.~A. 1993, \mnras, 265,
  588.

\bibitem[\protect\astroncite{{Dziembowski} \&
  {Pamyatnykh}}{2008}]{Dziembowski2008}
{Dziembowski}, W.~A. \& {Pamyatnykh}, A.~A. 2008, MNRAS, 385, 2061.

\bibitem[\protect\astroncite{{Gautschy} \& {Saio}}{1993}]{Gautschy1993}
{Gautschy}, A. \& {Saio}, H. 1993, \mnras, 262, 213.

\bibitem[\protect\astroncite{{Higl} \& {Weiss}}{2017}]{Higl2017}
{Higl}, J. \& {Weiss}, A. 2017, A\&A, 608, A62.

\bibitem[\protect\astroncite{{Hui-Bon-Hoa} \& {Vauclair}}{2018}]{HuiBonHoa2018}
{Hui-Bon-Hoa}, A. \& {Vauclair}, S. 2018, A\&A, 610, L15.

\bibitem[\protect\astroncite{{Iglesias} \& {Rogers}}{1991}]{Iglesias1991}
{Iglesias}, C.~A. \& {Rogers}, F.~J. 1991, ApJ, 371, L73.

\bibitem[\protect\astroncite{{Iglesias} \& {Rogers}}{1996}]{Iglesias1996}
{Iglesias}, C.~A. \& {Rogers}, F.~J. 1996, ApJ, 464, 943.

\bibitem[\protect\astroncite{{Iglesias} \emph{et~al.}}{1992}]{Iglesias1992}
{Iglesias}, C.~A., {Rogers}, F.~J., \& {Wilson}, B.~G. 1992, ApJ, 397, 717.

\bibitem[\protect\astroncite{{Kurucz}}{2004}]{Kurucz2004}
{Kurucz}, R. 2004, http:// kurucz.harvard.edu.

\bibitem[\protect\astroncite{{Lanz} \& {Hubeny}}{2007}]{Lanz2007}
{Lanz}, T. \& {Hubeny}, I. 2007, ApJ, 169, 83.

\bibitem[\protect\astroncite{{Lee} \& {Saio}}{1997}]{Lee1997}
{Lee}, U. \& {Saio}, H. 1997, ApJ, 491, 839.

\bibitem[\protect\astroncite{{Moskalik} \& {Dziembowski}}{1992}]{Moskalik1992}
{Moskalik}, P. \& {Dziembowski}, W.~A. 1992, A\&A, 256, L5.

\bibitem[\protect\astroncite{{Pamyatnykh}}{1999}]{Pamyatnykh1999}
{Pamyatnykh}, A.~A. 1999, Acta Astron., 49, 119.

\bibitem[\protect\astroncite{{Pamyatnykh} \emph{et~al.}}{2004}]{Pamyatnykh2004}
{Pamyatnykh}, A.~A., {Handler}, G., \& {Dziembowski}, W.~A. 2004, MNRAS, 350,
  1022.

\bibitem[\protect\astroncite{{Papics} \emph{et~al.}}{2017}]{Papics2017}
{Papics}, P.~I., {Tkachenko}, A., {Van Reeth}, T., \& {et al.} 2017, A\&A, 598,
  A74.

\bibitem[\protect\astroncite{{Paxton} \emph{et~al.}}{2011}]{Paxton2011}
{Paxton}, B., {Bildsten}, L., {Dotter}, A., \& {et al.} 2011, ApJS, 192, 3.

\bibitem[\protect\astroncite{{Paxton} \emph{et~al.}}{2013}]{Paxton2013}
{Paxton}, B., {Cantiello}, M., {Arras}, P., \& {et al.} 2013, ApJS, 208, 4.

\bibitem[\protect\astroncite{{Salmon} \emph{et~al.}}{2012}]{Salmon2012}
{Salmon}, S., {Montalb{\'a}n}, J., {Morel}, T., \& {et al.} 2012, MNRAS, 422,
  3460.

\bibitem[\protect\astroncite{{Seaton}}{1993}]{Seaton1993}
{Seaton}, M.~J. 1993, In \emph{Inside the Stars, IAU Coll. 137}, \emph{ASP
  Conference Series}, vol.~40, p. 222.

\bibitem[\protect\astroncite{{Seaton}}{2005}]{Seaton2005}
{Seaton}, M.~J. 2005, MNRAS, 362, L1.

\bibitem[\protect\astroncite{{Szewczuk} \&
  {Daszy{\'n}ska-Daszkiewicz}}{2018}]{Szewczuk2018}
{Szewczuk}, W. \& {Daszy{\'n}ska-Daszkiewicz}, J. 2018, MNRAS, 478, 2243.

\bibitem[\protect\astroncite{{Townsend}}{2003}]{Townsend2003}
{Townsend}, R.~H.~D. 2003, MNRAS, 340, 1020.

\bibitem[\protect\astroncite{{van Leeuwen}}{2007}]{vanLeeuwen2007}
{van Leeuwen}, F. 2007, A\&A, 474, 653.

\bibitem[\protect\astroncite{{Walczak} \emph{et~al.}}{2017}]{Walczak2017}
{Walczak}, P., {Daszy{\'n}ska-Daszkiewicz}, J., {Pamyatnykh}, A.~A., \& {et
  al.} 2017, In \emph{European Physical Journal Web of Conferences}, \emph{EDP
  Sciences}, vol. 152, p. 06005.

\end{thebibliography}

\end{document}